\documentclass[aps,prl,twocolumn,superscriptaddress,10pt]{revtex4-1}
\usepackage{graphicx,amsmath,amssymb,amsfonts}

\begin{document}

\title{Exciton-spin memory with a semiconductor quantum dot molecule}

\author{A. Boyer de la Giroday}
\affiliation{Toshiba Research Europe Limited, Cambridge Research Laboratory, 208 Science Park, Milton Road, Cambridge, CB4 0GZ, U. K.}
\affiliation{Cavendish Laboratory, Cambridge University, JJ Thomson Avenue, Cambridge, CB3 0HE, U. K.}
\author{N. Sk\"{o}ld}
\affiliation{Toshiba Research Europe Limited, Cambridge Research Laboratory, 208 Science Park, Milton Road, Cambridge, CB4 0GZ, U. K.}
\author{R.M. Stevenson}
\affiliation{Toshiba Research Europe Limited, Cambridge Research Laboratory, 208 Science Park, Milton Road, Cambridge, CB4 0GZ, U. K.}
\author{I. Farrer}
\affiliation{Cavendish Laboratory, Cambridge University, JJ Thomson Avenue, Cambridge, CB3 0HE, U. K.}
\author{D.A. Ritchie}
\affiliation{Cavendish Laboratory, Cambridge University, JJ Thomson Avenue, Cambridge, CB3 0HE, U. K.}
\author{A.J. Shields}
\email[]{\texttt{andrew.shields@crl.toshiba.co.uk}}
\affiliation{Toshiba Research Europe Limited, Cambridge Research Laboratory, 208 Science Park, Milton Road, Cambridge, CB4 0GZ, U. K.}
\date{\today}
\begin{abstract}
We report on a single photon and spin storage device based on a semiconductor quantum dot molecule. Optically excited single electron-hole pairs are trapped within the molecule and their recombination rate is electrically controlled over three orders of magnitude. Single photons are stored up to 1$\mu$s and read out on a sub-nanosecond timescale. Using resonant excitation, the circular polarisation of individual photons is transferred into the spin state of electron-hole pairs with a fidelity above 80$\%$, which does not degrade for storage times up to the 12.5ns repetition period of the experiment.
\end{abstract}
\maketitle
Further development of the emerging fields of quantum computing and quantum communication require quantum interfaces between optical ``flying'' and solid-state ``stationary'' qubits that enable reversible transfer, storage and manipulation of quantum information \cite{tanzilli}. Semiconductors quantum dots (QDs) constitute promising candidates as quantum interfaces between light and matter. Indeed, optical absorption of photons to create an electron-hole pair (or exciton) provides a natural mechanism to transfer quantum information from optical fields into the solid-state. Moreover, optical \cite{ramsey}, electrical \cite{boyer}, and optoelectrical \cite{zrenner,vasconcellos} coherent control of an exciton qubit in a QD have been successfully demonstrated. However, those approaches do not permit storage of quantum information, which limits their usefulness to the short lifetime of the exciton.

QD-based devices offering spin storage have been developed \cite{kroutvar,young,heiss} but rely on the storage of single carriers. Therefore, only pure spin states corresponding to the spin eigenstates can be stored. Indeed, for any other exciton spin state, the electron and the hole spins are entangled and loss of one of the carriers makes it impossible to recreate the original photon state, which prevents any extension of those schemes to a quantum memory.

Exciton storage has been demonstrated in ensembles of quantum dot molecules (QDMs) \cite{lundstrom}, quantum posts \cite{krenner1}, nanocrystals \cite{high}, natural QDs \cite{kraus} and coupled quantum wells \cite{rocke,zimmermann,winbow}. In those works the carriers were spatially separated for the storage operation. However, those works do not address the issue of transcribing and storing the polarisation of a photon into the spin state of the spatially-separated electron-hole pair, a necessary step towards a quantum memory for polarised light.

In this Letter, we present a novel approach which allows for initialisation and storage of the exciton spin state in a semiconductor QDM. We first demonstrate microsecond light storage at the single photon level. This is achieved by controlling the recombination rate of the electron-hole pair through electric-field dependent spatial separation in the QDM. We then extend our scheme to exciton-spin storage. Using a vertical magnetic field to lift the degeneracy of the exciton spin states, we show high-fidelity transfer and storage of the circular polarisation of photons into the exciton spin state under resonant optical excitation, with no observable degradation of the fidelity of the operation with time.

The device used consists of a $p$-$i$-$n$ heterostructure grown by molecular beam epitaxy. Two layers of InAs QDs separated by a thin GaAs layer were grown at the center of the intrinsic region made of a GaAs quantum well clad with a short period superlattice equivalent to Al$_{0.75}$Ga$_{0.25}$As on each side, which prevents tunneling of the carriers out of the QDM region when a large vertical electric field is applied. Vertical alignment of the QDs to form QDMs naturally arises from the strain field generated by the growth of the first layer. Doping extends into the superlattice and allows application of an electric field along the growth direction. This $p$-$i$-$n$ device is encased in a weak planar microcavity to enhance collection efficiency. We used standard photolithography and wet etching techniques to fabricate a diode with an area of 35$\times$60 $\mu$m$^{2}$. An opaque metallic film with $\mu$m-sized apertures allowed single QDMs to be addressed optically. All our experiments were performed on samples cooled to $<$10K in a liquid helium cryostat. A diode laser triggered by a pulse pattern generator of tunable period was used for above-band excitation. Resonant and quasi-resonant excitation was achieved using a wavelength-tunable Ti:Sapphire pulsed laser of linewidth $<30\mu eV$ operating at a fixed period of 12.5ns.

A schematic band diagram of the device is shown in Fig. 1(a). QDs from the top layer are strain-induced and therefore larger than QDs from the bottom layer. The electron (hole) energy states are consequently lower (higher) in the top QD than in the bottom QD. The exciton naturally relaxes into the top QD at zero electric field as it is the lowest energy configuration (``direct state''). The top QD is positioned on the p-doped side of the diode structure so that applying an electric field in reverse bias separates the hole energy levels of the top and bottom QDs further. On the other hand, the electron energy levels from both QDs are tuned into resonance at a field $F_{0}$ and the electron wavefunction hybridises over both QDs. Further increase of the electric field leads to further polarisation of the electron into the bottom QD, the exciton being eventually delocalised over both QDs forming the molecule (``indirect state'').

The energy of the exciton state as a function of electric field is accurately modelled using the direct and indirect states as basis states, which leads to the Hamiltonian \cite{stinaff}
\begin{equation}
\boldsymbol{H}=E_{0}+
\left( \begin{array}{cc}
0  & t           \\
t  & ed(F-F_{0}) \\
\end{array} \right),
\end{equation}
where $E_{0}$ is the energy of the exciton confined in the top QD at zero electric field, $F$ is the electric field applied to the structure, $t$ is the tunneling rate of the electron from the top to the bottom QD, $e$ is the charge of the electron, $d$ is the distance between the QDs, and $F_{0}$ is the field at which electron energy levels are tuned into resonance. The eigenstates of this Hamiltonian are the symmetric (low-energy) and antisymmetric (high-energy) combinations of the basis states \cite{dotyanti}. Calculated photoluminescence (PL) spectra of the emission from both eigenstates as a function of electric field is shown in Fig. 1(b), where the dotted lines correspond to the energy of the direct and indirect states. Coherent tunneling of the electron results in an anticrossing between the eigenstates. The dipole moment $ed$ of the exciton in its indirect state results in a linear Stark shift of the emission energy.

Photon and exciton-spin storage exploits the long lifetime of the exciton indirect state originating from the electron-hole spatial separation. Fig. 1(c) shows a schematic of the storage operation. At t=0, a laser pulse creates an electron-hole pair which relaxes into the QDM. The electric field is such that the exciton is in an indirect state characterised by a long lifetime which prevents recombination of the electron and the hole during a storage time $\Delta t$. At $t=\Delta t$, a sub-nanosecond modulation of the electric field is applied to bring the exciton from the indirect to the direct state so that readout of the emission occurs naturally through recombination.
\begin{figure}
	  \includegraphics[width=1\columnwidth,keepaspectratio]{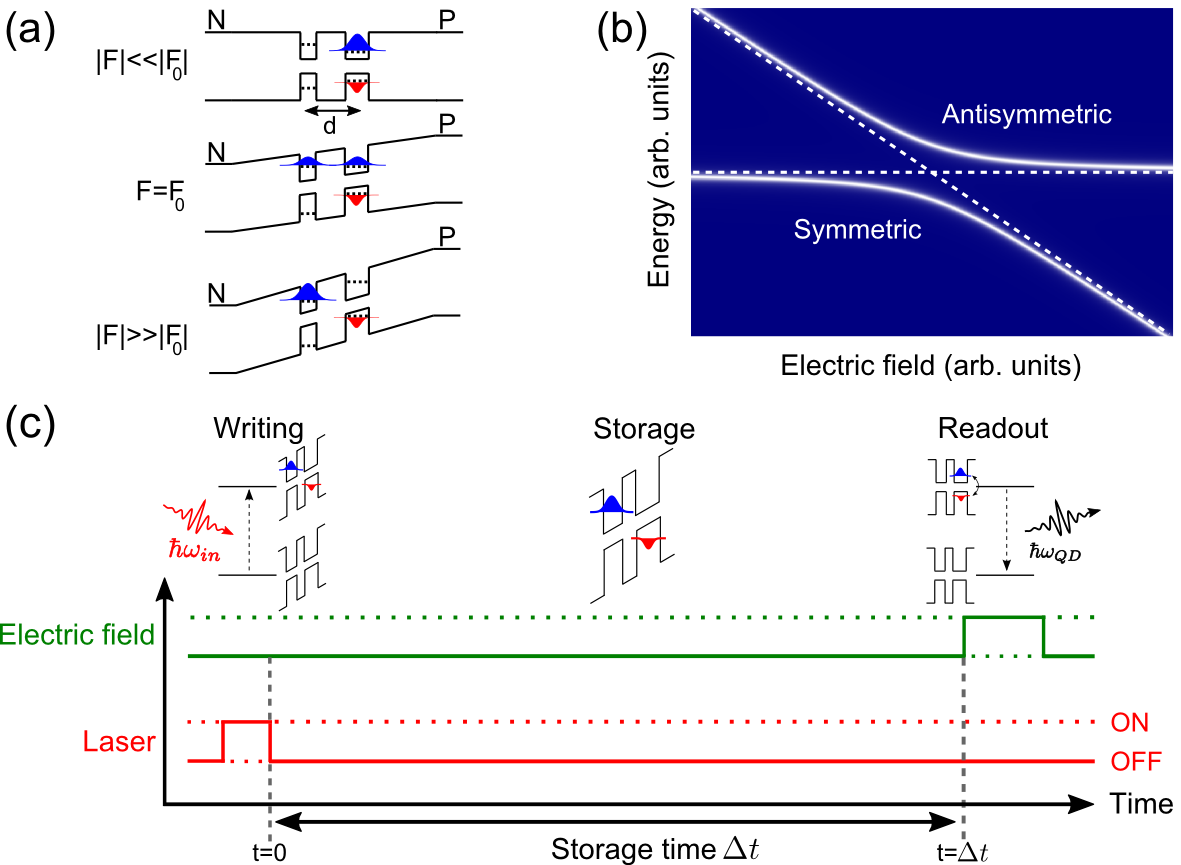}
	\caption{(colour online). (a) Schematic of the band-structure of the QDM at various electric fields. The electron and hole wavefunctions for the low-energy eigenstate are illustrated respectively in blue and red. (b) Calculated PL spectra of the emission from the QDM eigenstates as a function of electric field. The dotted lines correspond to the energies of the direct and indirect states as a function of electric field. (c) Operating principle behind photon and exciton-spin storage.}
	\label{fig1}
\end{figure}

PL emission from the symmetric exciton state of the QDM used for photon storage is shown in Fig. 2(a) as a function of electric field. Fitting using Eq. (1) gives values of $d$=8.2$\pm$0.05nm and $t$=7.9$\pm$0.05meV. The high-energy antisymmetric states of the QDMs that we studied were either not visible (as in Fig. 2(a)) or of weak intensity and therefore not suitable for storage experiments. We attribute this to effective thermalization of the exciton to the low-energy symmetric state. Missing spectral lines from QDMs due to thermalization effects have also been reported in other studies \cite{scheibner2}. The exciton lifetime as a function of electric field measured using standard time-resolved spectroscopy (``direct measurement'') is shown in Fig. 2(b). A rapid increase in lifetime is observed as the electron wavefunction progressively polarises from the top to the bottom QD.

Using the method described in Fig. 1(c), we successfully stored the exciton for storage times up to $1\mu s$, corresponding to more than 1000 times the usual lifetime of an exciton in a single QD. The results are shown in Fig. 2(c) for storage times of 0.2, 0.4, 0.6, 0.8, and 1$\mu$s. The intensity $I$ of the readout emission decays with storage time $\Delta t$ due to the finite lifetime $\tau_{s}$ of the indirect state. This decay is given by $I(\Delta t)=I_{0}e^{-\frac{\Delta t}{\tau_{s}}}$, where $I_{0}$ is the intensity measured at t=0. The lifetime of the storage state can therefore be measured indirectly by fitting the decay to an exponential. This was done at different electric fields and the corresponding measurements are shown in Fig. 2(b) and compared to the measurements obtained by standard time-resolved spectroscopy. Agreement between both methods confirms the storage of the exciton in its indirect state. The maximum storage time was limited only by the pulse pattern generator used to trigger the laser pulse, which had a maximum operating period of 1.1$\mu$s. Storage times of several seconds can be expected using our scheme and engineering QDMs so that longer lifetimes can be achieved \cite{lundstrom}.
\begin{figure}
	  \includegraphics[width=1\columnwidth,keepaspectratio]{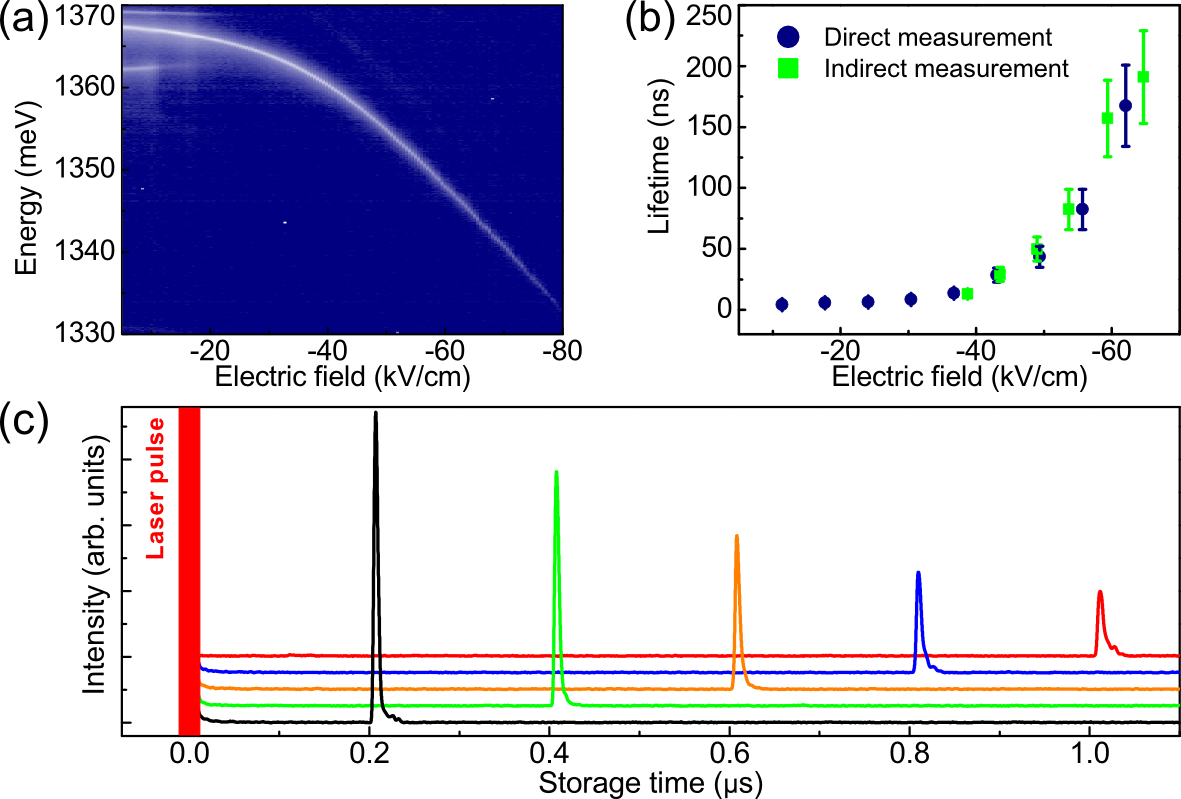}
	\caption{(colour online). (a) Electric-field dependence of the PL emission from the symmetric exciton state used for photon storage. (b) Exciton lifetime as a function of electric field. (c) Readout signal after storage in an indirect state, at an electric field of -77kV/cm, for storage times up to $1\mu s$.}
	\label{fig2}
\end{figure}

To store exciton-spins in a single QDM using our operating scheme, both resonant optical excitation and a vertical magnetic field are required. Resonant excitation will allow for high-fidelity transfer of the polarisation of a photon into the exciton spin state \cite{boyer} while a vertical magnetic field will lift the degeneracy of the exciton spin states and define two Zeeman eigenstates with circular polarisations and long spin lifetimes \cite{kroutvar,young}.

In Fig. 3(a), PL emission from the QDM used for exciton-spin storage is shown as a function of electric field. Fittings using Eq. (1) provides values of d=8.15$\pm$0.05nm and t=6$\pm$0.5meV. We also performed PL excitation spectroscopy at various electric fields and two clear resonances were found. Their energies are plotted as a function of electric field in Fig. 3(a). The first resonance was found at 36.7$\pm$0.6meV from the exciton energy. A similar resonance at the same energy was found as well for other QDMs, indicating resonant excitation through the creation of a 1-LO phonon in GaAs, the expected energy of which is $\approx$36.4meV \cite{strauch}. The second resonance appears at 55.2$\pm$2.1meV and is attributed to the $p$-shell of the exciton state. Consistent with this assignment, we found the position of this resonance varied from QDM to QDM. The lifetime as a function of electric field was measured using standard time-resolved spectroscopy and results are shown in Fig. 3(b). The lifetime levels out around 1$\mu$s for this particular QDM, which is related to the minimum overlap of the electron-hole wavefunctions reached in the indirect state.

The behaviour of our QDM under a vertical magnetic field was also studied. The energy of the exciton spin states as a function of magnetic field $B$ \cite{walck} is given by $E(B)=E_{0}+\gamma_{1}B+\gamma_{2}B^{2}+\ldots$, where $|\gamma_{1}|=g\mu_{B}/2$ provides a direct measurement of the exciton g-factor $g$ ($\mu_{B}$ is the Bohr magneton and the sign of $\gamma_{1}$ is given by the angular momentum $\pm$1 of the spin states), $\gamma_{2}$ is the diamagnetic coefficient, and where higher order terms can be neglected for the fields at which we are working ($B<4T$). Measurements of the exciton g-factor and the diamagnetic coefficient $\gamma_{2}$ as a function of electric field are shown in Fig. 3(c) and 3(d). No electric field dependence of the g-factor is found for electron tunneling, as electron g-factors are similar in bulk GaAs (-0.44) and InAs QD (-0.6) \cite{dotygfactor}. However, a strong dependence of the diamagnetic coefficient as a function of electric field is found and an order of magnitude difference is measured between the direct and indirect states. The diamagnetic coefficient is a measure of the in-plane electron-hole separation, which in turns depends on the lateral confinement and Coulomb interaction \cite{walck}. When the electron tunnels from the top to the bottom QD, the lateral confinement does not change significantly since both QDs have similar sizes. The increase in $\gamma_{2}$ in the indirect state therefore provides a measure of the reduced Coulomb interaction. $\gamma_{2}$ saturates at high electric field in the indirect state, where the electron-hole pair is fully spatially separated.
\begin{figure}
	  \includegraphics[width=1\columnwidth,keepaspectratio]{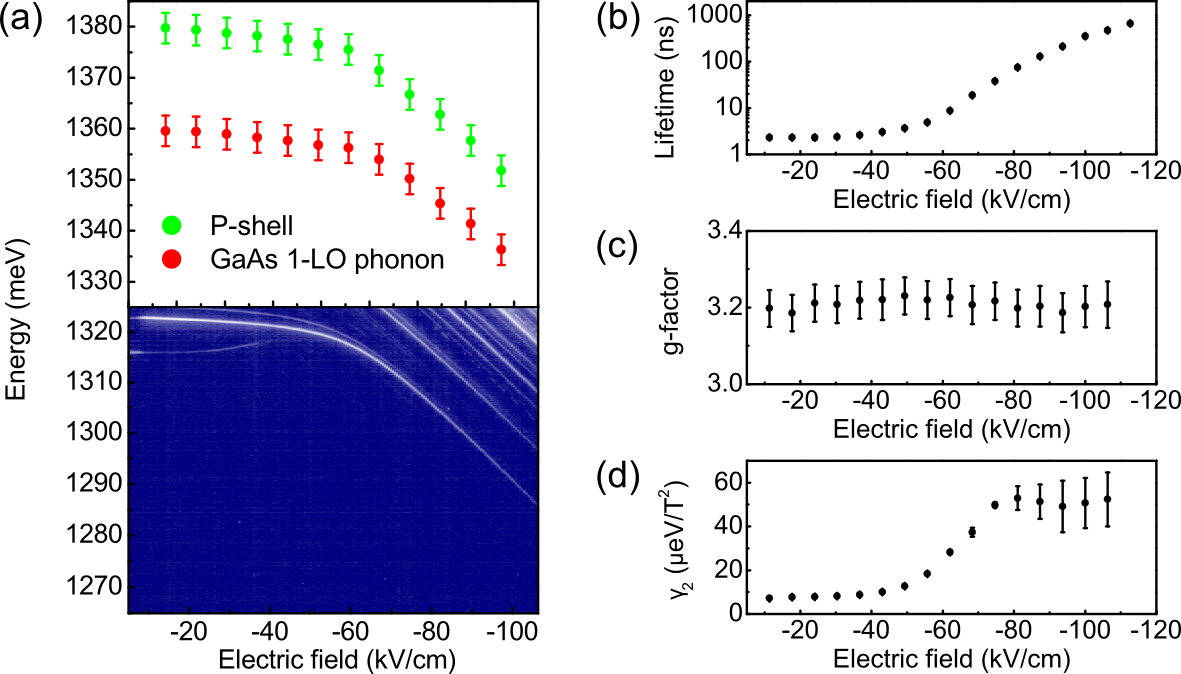}
	\caption{(colour online).  (a) PL spectrum of the emission of the exciton used for exciton-spin storage as a function of electric field along with the energies of the two resonances observed. (b) Exciton lifetime as a function of electric field. (c) and (d) g-factor and diamagnetic coefficient as a function of electric field. Data presented in (a) and (b) were taken at $B$=0T. Each data point presented in (c) and (d) was obtained from measurements of the energy dependence of the Zeeman states with magnetic field in the range 0T$<B<$4T.}
	\label{fig3}
\end{figure}

Circular spin eigenstates are created by applying a vertical magnetic field such that $g\mu_{B}B>>s$, where $s$ is the fine structure splitting of the exciton eigenstates. For our QDM, $s$ in the direct state at zero electric field is 76.5$\pm$0.1$\mu$eV and circular eigenstates are therefore created for $B>2T$. Exciting resonantly into the QDM while applying a vertical magnetic field allows for high-fidelity transfer of the photon circular polarisation into the exciton spin state. The fidelity of this transfer is defined as the degree of polarisation of the exciton emission $F=(I_{co}-I_{cross})/(I_{co}+I_{cross})$, where $I_{co}$ ($I_{cross}$) is the intensity from the Zeeman exciton state co(cross)-polarised with the excitation polarisation.

Fig. 4(a) shows the fidelity as a function of storage time for left- and right-hand circular pump polarisations and for three excitation schemes: $p$-shell, phonon and $s$-shell. Discrimination between the laser signal and the exciton emission when exciting into the $s$-shell was achieved by taking advantage of the large Stark-shift between the energy of the indirect ``write'' state and the direct ``read-out'' state. The storage time was limited only by the 12.5ns period of the laser used for the experiment. Moreover, ringing in the electrical signal prevented measurements at some storage times, as a change in electric field shifts the exciton energy out of the excitation or detection window through the Stark effect. This effect was particularly important in the case of $s$-shell excitation, where the resonance is the narrowest.
\begin{figure}
	  \includegraphics[width=1\columnwidth,keepaspectratio]{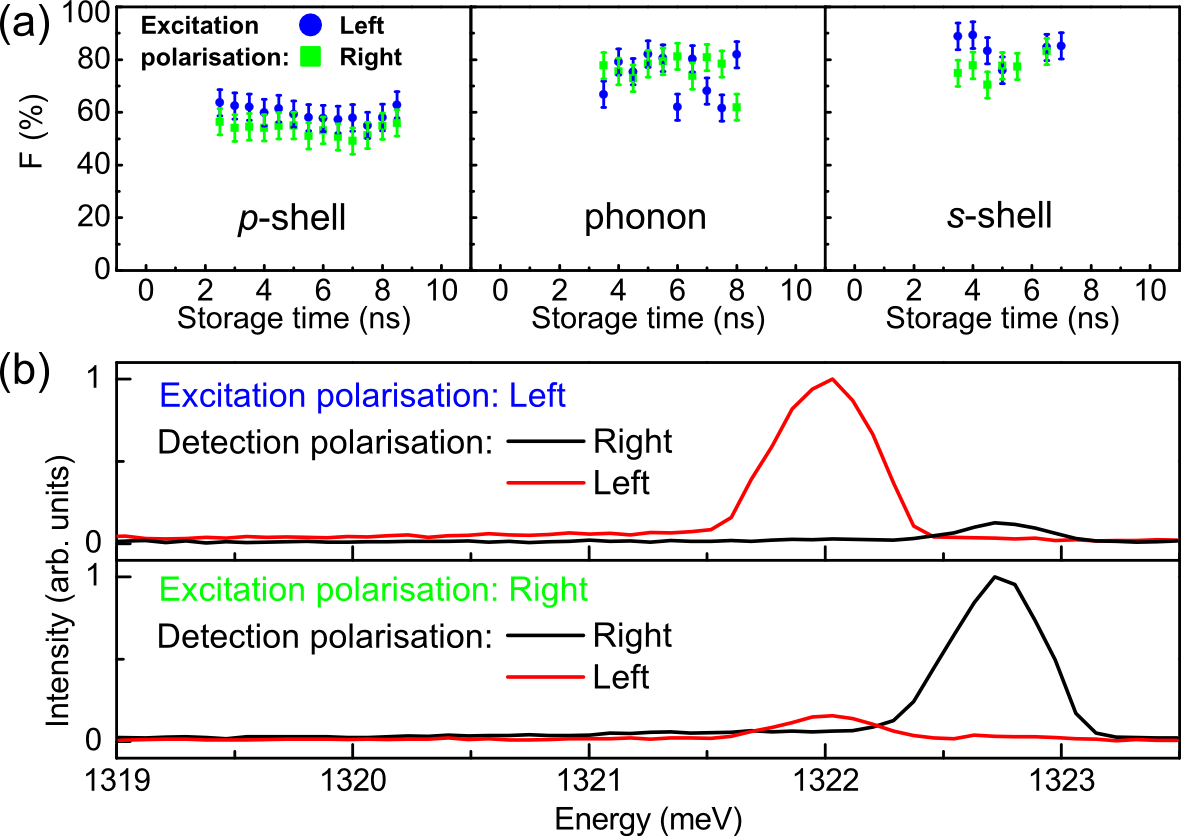}
	\caption{(colour online). (a) Fidelity of the exciton-spin memory operation as a function of storage time for different excitation schemes and polarisations. The data was taken at $B$=4T (similar results were obtained for 2T$<B<$4T). (b) Typical spectra obtained after a storage time of 5ns when exciting in the phonon resonance and at $B$=4T.}
	\label{fig4}
\end{figure}

The fidelity does not degrade over the 12.5ns repetition period of the experiment, suggesting that the spin relaxation time is substantially longer than the storage time. Average fidelities over time and over both pump polarisations are 57$\pm$5$\%$ for $p$-shell excitation, 75$\pm$7$\%$ for phonon excitation, and 81$\pm$6$\%$ for $s$-shell excitation. No spin memory effect was observed when exciting in the continuum states, which we attribute to spin relaxation as the charge carriers thermalize down to exciton ground state. The fidelity is usually found to be slightly higher when exciting with left-hand circularly polarised light (particularly when using $p$-shell excitation), which is attributed to thermal relaxation from the right-polarised (high-energy, see Fig. 4(b)) Zeeman eigenstate down to the left-polarised (low-energy) Zeeman eigenstate. Fig. 4(b) illustrates the fidelity of the stored emission obtained under phonon excitation. Comparable results were obtained when measuring the fidelity of the ``write'' state (without storage operation). As a consequence, the storage operation is found to have no observable effect on the fidelity and the limitation arises from the ``write'' operation, which may be improved by using wave plates calibrated at the QDM energy.

In conclusion, we have successfully stored single photons up to 1$\mu$s and exciton spins with a fidelity above $80\%$ in a single QDM. No decay in the fidelity was apparent within the period of the excitation laser of 12.5ns. Increasing the barrier thickness of the QDM, which would lead to longer lifetimes in the indirect state, and using a laser with a tunable period, we anticipate that spin storage in a single QDM is achievable for times in excess of 1ms. Moreover, it is possible to extend the presented exciton spin memory to a quantum memory and store a superposition of the spin eigenstates if these can be made near degenerate. The energy separation of the eigenstates is limited by the exciton fine structure splitting and studies of single QDs have shown that it can be reduced to a level where the precession of a superposition state can be recorded with standard time-resolved spectroscopy \cite{stevenson}. This offers new possibilities to study and improve the coherence of an exciton spin state, which is expected to be limited by hyperfine interaction between the electron spin and the nuclear spins \cite{hanson}. Eventually, several QDM quantum memories could be combined on a single chip. Recent advances in the field of solid-state cavity quantum electrodynamics makes it likely that such a system would be scalable \cite{obrien}.
\begin{acknowledgments}
We thank the EU for partial support through the FP7 FET Q-ESSENCE Integrated Project and the Spin-Optronics ITN Project. We thank K. Cooper for help with the processing. A.B.d.l.G would also like to thank EPSRC and TREL for financial support.
\end{acknowledgments}

\end{document}